\documentclass[epjCONF,columns]{svjour} %for 2 columns format
\pdfoutput=1
\def\Mbb   {\ensuremath{M(\mathrm{jj})}}
\def\BDT {{\tt BDT}}
\def\Zj  {\ensuremath{\mathrm{Z + jets}}}
\def\Wj  {\ensuremath{\mathrm{W + jets}}}

\def\ttbar{\ensuremath{t \bar{t}}}
\def\ZnnH  {\ensuremath{\mathrm{Z}(\to \nu \nu)\mathrm{H}}}

\def\ZmmH  {\ensuremath{\mathrm{Z}(\to \mu \mu)\mathrm{H}}}
\def\ZeeH  {\ensuremath{\mathrm{Z}(\to e e)\mathrm{H}}}

\def\WmnH  {\ensuremath{\mathrm{W}(\to \mu \nu)\mathrm{H}}}
\def\WenH  {\ensuremath{\mathrm{W}(\to e \nu)\mathrm{H}}}

\def\WJ  {\ensuremath{\mathrm{W + jets}}}
\def\HBB   {\ensuremath{\mathrm{H}\to b\bar{b}}}

\def\Vudscg   {\ensuremath{\mathrm{V+udscg}}}
\def\Wudscg   {\ensuremath{\mathrm{W+udscg}}}

\def\Zudscg   {\ensuremath{\mathrm{Z+udscg}}}

\def\Wbb   {\ensuremath{\mathrm{W\to b\bar{b}}}}
\def\Zbb   {\ensuremath{\mathrm{Z\to b\bar{b}}}}

\def\Mjj   {\ensuremath{M(\mathrm{jj})}}

\def\dEtaJJ {\ensuremath{\Delta \eta(\mathrm{J1,J2})}}
\def\dphiVH {\ensuremath{\Delta\phi(\mathrm{V,H})}}

\def\dphiMJ {\ensuremath{\Delta\phi(\mathrm{pfMET,J})}}

\def\ptV {\ensuremath{p_{\mathrm{T}}(\mathrm{V})}}

\def\Naj {\ensuremath{N_{\mathrm{aj}}}}
\def\Nal {\ensuremath{N_{\mathrm{al}}}}

\def\Bexp {\ensuremath{B_{\mathrm{exp}}}}

\def\Nobs {\ensuremath{N_{\mathrm{obs}}}}

\usepackage{graphics}
\usepackage[varg]{txfonts} % Times fonts
\usepackage[latin1]{inputenc}
\session-title{Conference Title, to be filled}
\begin{document}
\title{Search for a Standard Model Higgs boson decaying to b quarks and produced in
association with Z/W bosons with the CMS detector}
\author{Michele De Gruttola\inst{1}\fnmsep\inst{2}\fnmsep\thanks{\email{michele.de.gruttola@cern.ch}} }
\institute{University of Florida, Institute for High  Energy Physics
  and Astrophysics, Department of Physics, Gainesville, FL USA
  \and Fermi National Accelerator Laboratory, Batavia, IL USA}
\abstract{
A search for the standard model Higgs boson is performed in a data
sample corresponding to an integrated luminosity of 1.1 fb$^{-1}$,
recorded by the CMS detector~\cite{CMS} in proton-proton collisions at the LHC
with a 7 TeV center-of-mass energy. The following modes are studied: W($\mu \nu$)H, W(e$\nu$)H, Z($\mu \mu$)H, Z(ee)H and Z($\nu \nu$)H, with the Higgs decaying to bb pairs. 95\% C.L. upper limits on the VH production cross section are derived for a Higgs mass between 110 and 135 GeV. The expected (observed) upper limit at 115 GeV is found to be 5.7 (8.3) times the standard model expectation.
} %end of abstract
\maketitle
\section{Introduction}
\label{intro}
The search for the Higgs boson~\cite{ref:Higgs1}  is currently one of
the most important undertakings of experimental particle physics.

At the LHC the main Higgs production mechanism is direct production through gluon fusion,
with a cross section of $\sim 17 \cdot 10^3$~fb for a Higgs mass $m_H= 120$~GeV~\cite{LHCXSWG}.
However, in this production mode, the detection of the $H\to b\bar b$ decay
is rendered nearly impossible due to overwhelming QCD di-jet production.
The same holds true for the next most copious production mode, through
vector-boson fusion, with a cross section of $\sim1,300$~fb. Instead
we consider processes in which the Higgs is produced in association with a vector
boson which have cross sections
of $\sim660$ and $\sim360$~fb for $WH$ and $ZH$ respectively. Even if
the resulting sensitivity of the $H\to b\bar b$ decay is less
than other final states such as $H\to\gamma\gamma$ and $H\to\tau\tau$
for example, it is paramount to search for the Higgs in
these modes given that the observation of the $H\to b\bar b$ decay is
key to determine the nature of this particle, if and when discovered.

We summarize a search for the standard model Higgs
boson in the $pp\to VH$ production mode with the CMS detector. The analysis
is performed in a data sample corresponding to an integrated luminosity of
 $1.1$~fb$^{-1}$, collected by the CMS experiment at a
7~TeV center-of-mass energy. The following final states are included:
$W(\mu\nu)H$, $W(e\nu)H$, $Z(\mu\mu)H$, $Z(ee)H$ and $Z(\nu\nu)H$
 --all with the Higgs decaying to $b\bar{b}$ pairs. 

Backgrounds arise from production of W and Z bosons associated with
jets (all flavors), singly (ST) and pair-produced top quarks, and
di-bosons (VV).  
Simulated samples of all backgrounds are used to provide
guidance in the analysis optimization, and an initial evaluation 
of their contributions in the search region.
For the main backgrounds, high-purity control regions are used to
estimate their contribution in the signal region.

An optimization of the event selection, that depends on the Higgs mass, is performed, and 95\%
C.L. upper limits on the  $pp\to VH$ production cross section are
obtained for Higgs masses between
110-135 GeV. These limits are based on the observed event count and background estimate in signal 
regions defined in either the invariant mass distribution of $H\to
b\bar b$ candidates (``\Mbb\, or cut-and-count analysis''), or in the output
discriminant of a boosted decision tree algorithm (``\BDT\ analysis'')\cite{BDT}. The latter enhances
the statistical power of the analysis by making full use of correlations between discriminating 
variables in signal and background events.

For lack of space we will present here only tables and plots for the 115 GeV mass hypothesis, while only the final limits plots will contains all mass range search.   

%A detailed description of the CMS detector can be found
%elsewhere~\cite{CMS}. The key components of the detector include a
%silicon pixel and a silicon strip tracker, immersed in a 3.8~T
%solenoidal magnetic field, which are used to measure the momenta of
%charged particles. The tracker, which covers the pseudorapidity range
%$|\eta| < 2.5$, is surrounded by a crystal electromagnetic
%calorimeter (ECAL) and a brass-scintillator hadronic calorimeter
%(HCAL). The ECAL and HCAL extend to a pseudorapidity range of $|\eta|
%< 3.0$. A quartz fiber Cherenkov forward detector (HF) extends the
%calorimetric coverage to $|\eta| < 5.0$. The outermost component of
%the CMS detector is the muon system consisting of gas detectors
%placed in the iron return yoke to measure the momentum of muons traversing through the detector.

\section{Event selection}
\label{sec:eventSelection}

Candidate $W( \to \ell \nu)$ decays are identified by requiring
the presence of a single, isolated, lepton and additional missing
transverse energy(MET). Muons (electrons) are required to have a $p_t$ above
20 (30)~GeV. Candidate $Z \to \ell \ell$ decays are reconstructed by combining 
isolated, opposite charge pairs of electrons and muons and requiring the dilepton invariant 
mass to satisfy $75<$m$_{ \ell \ell}<105$GeV.  For Z candidates the electron $p_t$ is
lowered to 20~GeV.
The identification of $Z \to \nu\nu$ decays requires MET~$>160$GeV (the high threshold dictated by the trigger).

The reconstruction of the $H\to b\bar b$ decay is made by
requiring the presence of two central ($|\eta|<2.5$)
jets, above a minimum $p_t$ threshold and b-tagged. If more than two such jets are
found in the event, the pair with the highest sum of the b-tag outputs
for the two jets is chosen (except for the WH analyses, in which
the $t\bar t$ background is larger, where the pair of jets with
highest total $p_t$ is chosen). These combinations are found to yield
         higher efficiency and rejection of wrong combinations in signal
         events, as opposed to simply selecting the two highest $p_t$ jets
         in the event. After b-tagging the fraction of $H\to b\bar b$
         candidates that contain the two b-jets from the Higgs decay
         is near unity.
  
After b-tagging, the background from V+jets and di-bosons is reduced significantly and
becomes dominated by the sub-processes where the two jets originate from real b-quarks.
Events with additional jets (\Naj\ ) or additional leptons (\Nal\ ) are rejected to
further reduce backgrounds from $t\bar{t}$ and WZ.

The topology of VH production is such that the W/Z
and the Higgs recoil away from each other with significant $p_t$. 
Cuts on the azimuthal opening angle between the vector 
boson and the reconstructed momenta of the Higgs candidate, \dphiVH,
on the $p_t$ of the V-boson and on the b-tagged dijet pair
achieve significant rejection for most background processes and
improve the analysis reach.

For the $Z \to \nu \nu$ channel, QCD backgrounds are further reduced
by a factor of $\sim 30$ when requiring that the MET does not originate from
mismeasured jets. 

The training of the \BDT\ is done with simulated samples for signal and background
that pass a looser event selection relative to the \Mbb\ analyses. 
 Several input variables were 
chosen by iterative optimization. These include the di-jet invariant
mass and momentum: \Mjj\ and $p_{tjj}$ ,
the V transverse momentum \ptV\ , the b-tag value for each of the two
jets, the azimuthal
angle between the V and the dijets, \dphiVH\ , and the pseudorapidity 
separation between the two jets, \dEtaJJ\ .
The \BDT\
 analysis was expected to improve the sensitivity with respect
to the \Mbb\ analysis by about 10\%
 in every channel.

\section{Control regions}
Appropriate control regions that are orthogonal to the signal region are 
identified in data and used to adjust Monte Carlo estimates for 
the most important background processes: \Wj\ and \Zj\ (with
light and heavy-flavor jets), $t\bar{t}$ and QCD
multijet and heavy-quark production.  Different control regions are
found for each of the different search channels by changing the event
selection in a way that enriches the content of each specific
background. For all cases, control regions with purity ranging 
from about 20\% to nearly 100\% have been successfully found. Discrepancies between the expected and 
observed yields in the data in these control regions are used to
obtain a scale factor by which the estimates from the simulation are
adjusted. The background from these sources in the signal region are then estimated from
the adjusted simulation samples, taking into account the associated
systematic uncertainty. The precise construction of all the control regions is involved and outside the 
scope of this summary. The procedures applied include, for example: reversing the 
b-tagging requirements to enhance \Wj\ and \Zj\ with light-flavor jets; enforcing a tighter
b-tag requirement and requiring extra jets to enhance $t\bar t$ and requiring low ``boost'' in order
to enhance $V \to b b$ over $t\bar t$ . Table~\ref{tab:MCSF} lists the control regions
and the corresponding purities and scale factors obtained.

\begin{table}[tbp]
\caption{Purity and scale factors (Data/MC) derived from background
  enriched control regions (CR), as described in the text. The scale
  factors for $W \to \mu \nu H $and $ W \to e \nu H $ were averaged together, and the same
  was done for $Z \to \mu \mu H$ and $Z \to ee H$.}
\label{tab:MCSF}
\begin{center}
\begin{footnotesize}
\begin{tabular}{ccccccc}\hline\hline
 &    \multicolumn{2}{c} {$W \to \ell \nu H$}    & \multicolumn{2}{c} {$Z \to  \ell \ell H$}         \\ 
CR  & Purity  & SF     &  Purity  & SF \\ \hline
\Vudscg           &  79.4\% & $0.84\pm 0.10$ & 92.8\% &  $0.88\pm 0.02$ \\ \hline
$t\bar t$          & 85.8\% & $1.01\pm 0.11$ &  97.5\% & $0.99\pm 0.05$ \\ \hline
V+$b\bar b$  &  20.2\% & $1.40\pm 0.29$ & 81.6\% & $1.16\pm 0.08 $   \\ \hline\hline
\end{tabular}
\end{footnotesize}
\end{center}
\end{table}

The $Z \to \nu \nu H$ channel is unique among the five modes analyzed, in that
it does not include charged leptons.  An important check is to 
compare the observed pfMET distribution with the predicted distribution
from simulation.  To accomplish this, muons are removed from the 
$Z \to \mu \mu J$ data sample. 
Reasonably pure samples of $t\bar{t}$ and \WJ\ events can be obtained by
requiring at least one additional isolated lepton in the event, and then
either requiring (for $t\bar{t}$) or vetoing (for \WJ) b-jets. 
Table~\ref{tab:MCSF_Zinv} 
lists the control regions and the corresponding purities and scale
factors obtained. 

The QCD background in the signal region is also estimated from data
using control regions of high and low values of two uncorrelated
variables with significant discriminating power towards QCD events.
One is the angle between the missing
energy vector and the closest jet in azimuth, \dphiMJ\, and the other
is the sum of the CSV values of the two b-tagged jets. The signal
region is at high values of both discriminants, while QCD populates
regions with low values of either. The method predicts a negligible
contamination of this background.

\begin{table}[tbp]
\caption{$Z_{inv}$ Purity and scale factors (SF, Data/MC) derived 
from background enriched control regions, as described in the text.}
\label{tab:MCSF_Zinv}
\begin{center}
\begin{tabular}{ccc}\hline\hline
CR  &  Purity         & SF   \\ \hline\hline
\Zudscg           &    92.4\%        &   $0.97\pm 0.06$     \\ \hline
\Wudscg           &   94.1\%        &   $0.92\pm 0.05$     \\ \hline
$Z\to bb$               &    44.4\%        &   $1.00\pm 0.30$     \\ \hline
$t\bar t$  &    89.9\%        &   $0.91\pm 0.09$     \\\hline
\end{tabular}
\end{center}
\end{table}

\section{Systematics}
The following systematic uncertainties on the expected signal and background
yields affect the upper limit. The values listed are an approximation of
what is actually used in the limit calculation.

The total uncertainty on the signal prediction is taken to be $26\%$ and 
$28\%$ for ZH and WH production, respectively.  Background uncertainties 
range from $12\%$ to $20\%$ depending on mode and mass point.

Experimental sources of systematics are the b-tag efficiency
($\sim$10\%), the jet energy resolution ($\sim$10\%) and scale ($\sim$1\%) uncertainty, the machine luminosity ($\sim$4.5\%), the trigger efficiency ($\sim$2\%).   The signal cross section is affected by electroweak corrections for a boost of $\sim 150$~GeV
    are $5\%$ for ZH and $10\%$ for WH, and  QCD correction,  relevant  in the comparison  NNLO vs. NLO, where an uncertainty of 
  $10\%$ for both ZH and WH is estimated. 

\section{Results}
\begin{figure}[htbp]
\begin{center}
%\resizebox{0.75\columnwidth}{!}{  \includegraphics{figures/CCWm_Aug18.pdf} }
  \resizebox{0.62\columnwidth}{!} { \includegraphics{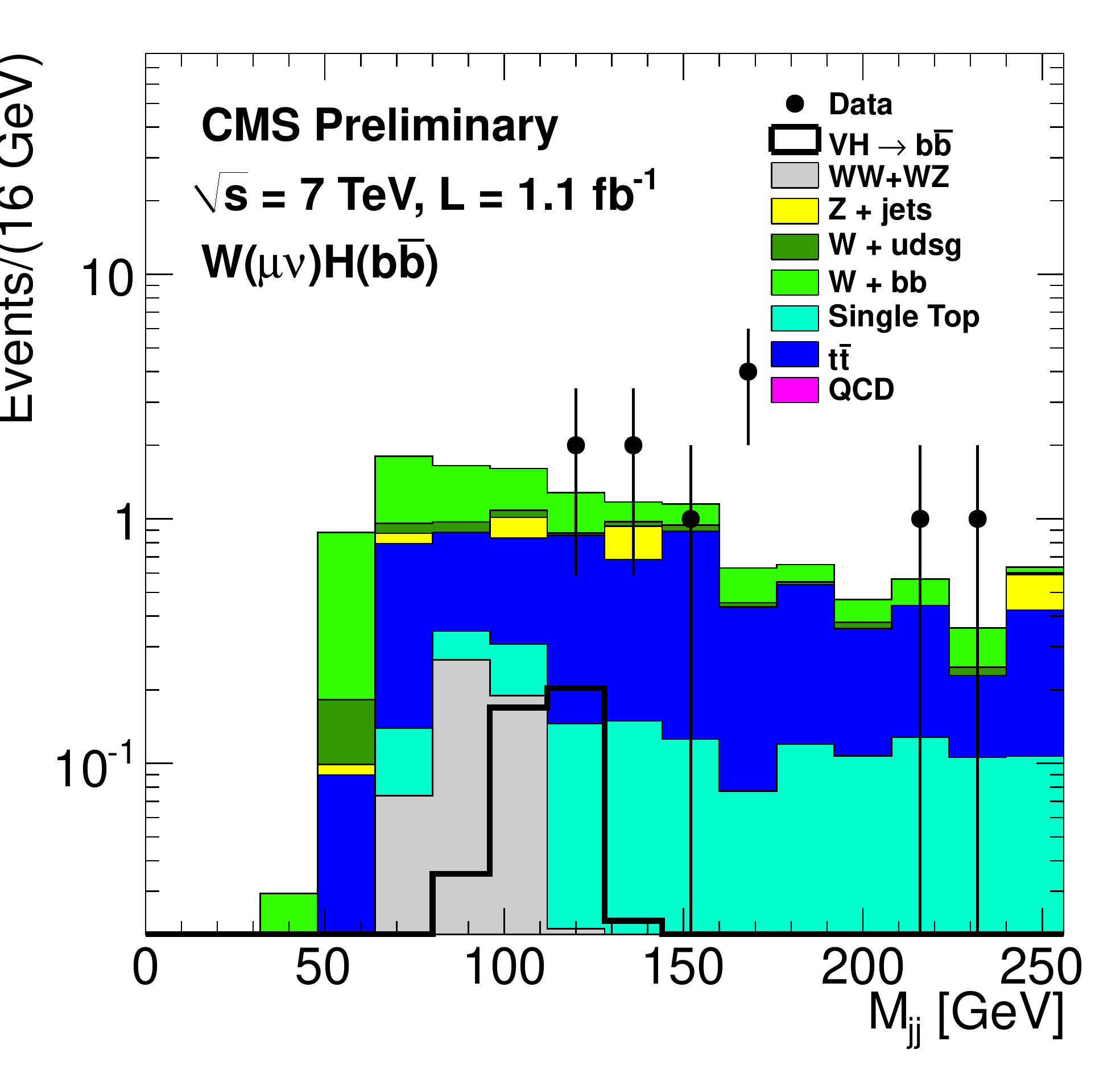} }
 \resizebox{0.62\columnwidth}{!}{  \includegraphics{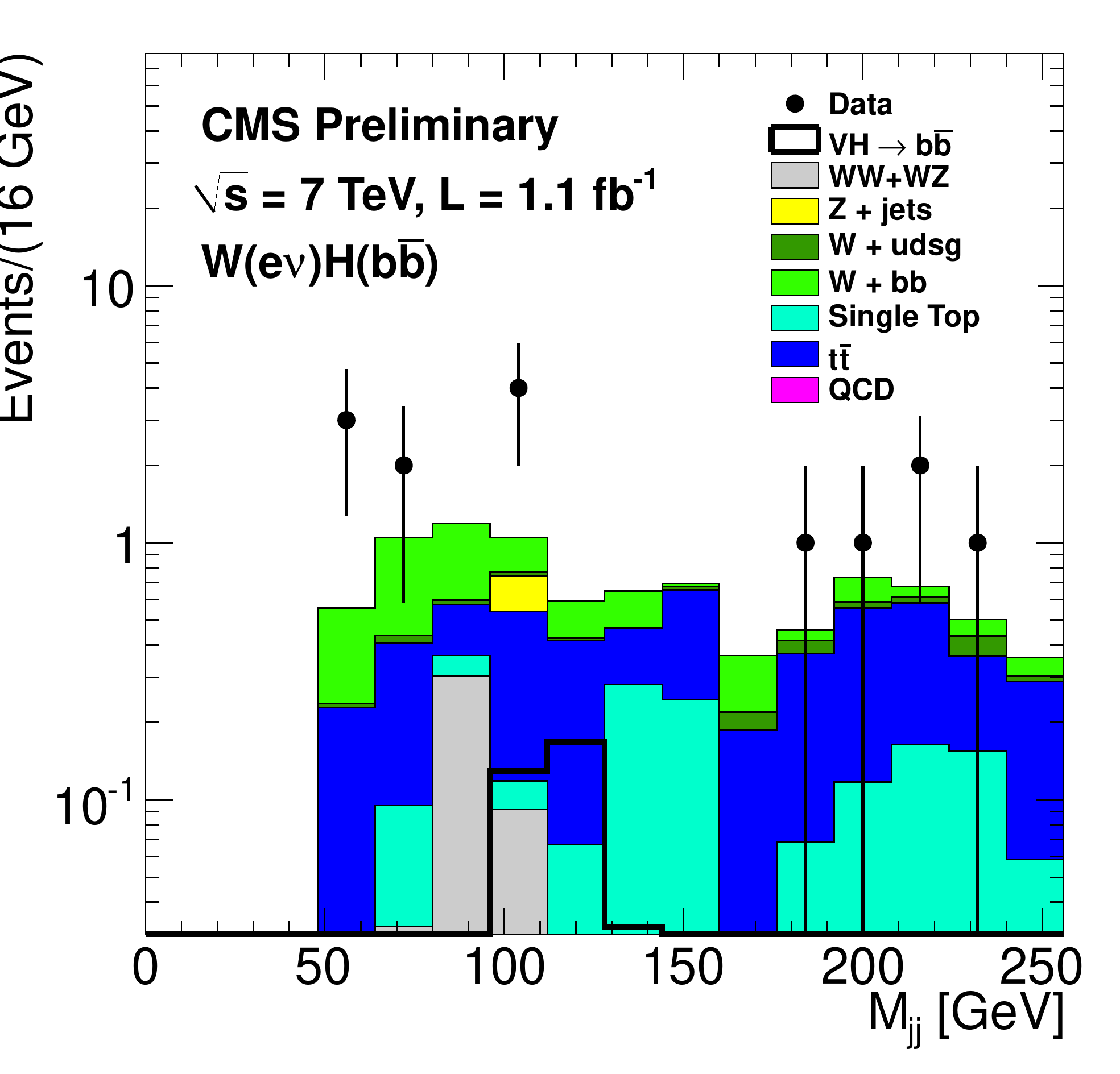}}
  \resizebox{0.62\columnwidth}{!}{  \includegraphics{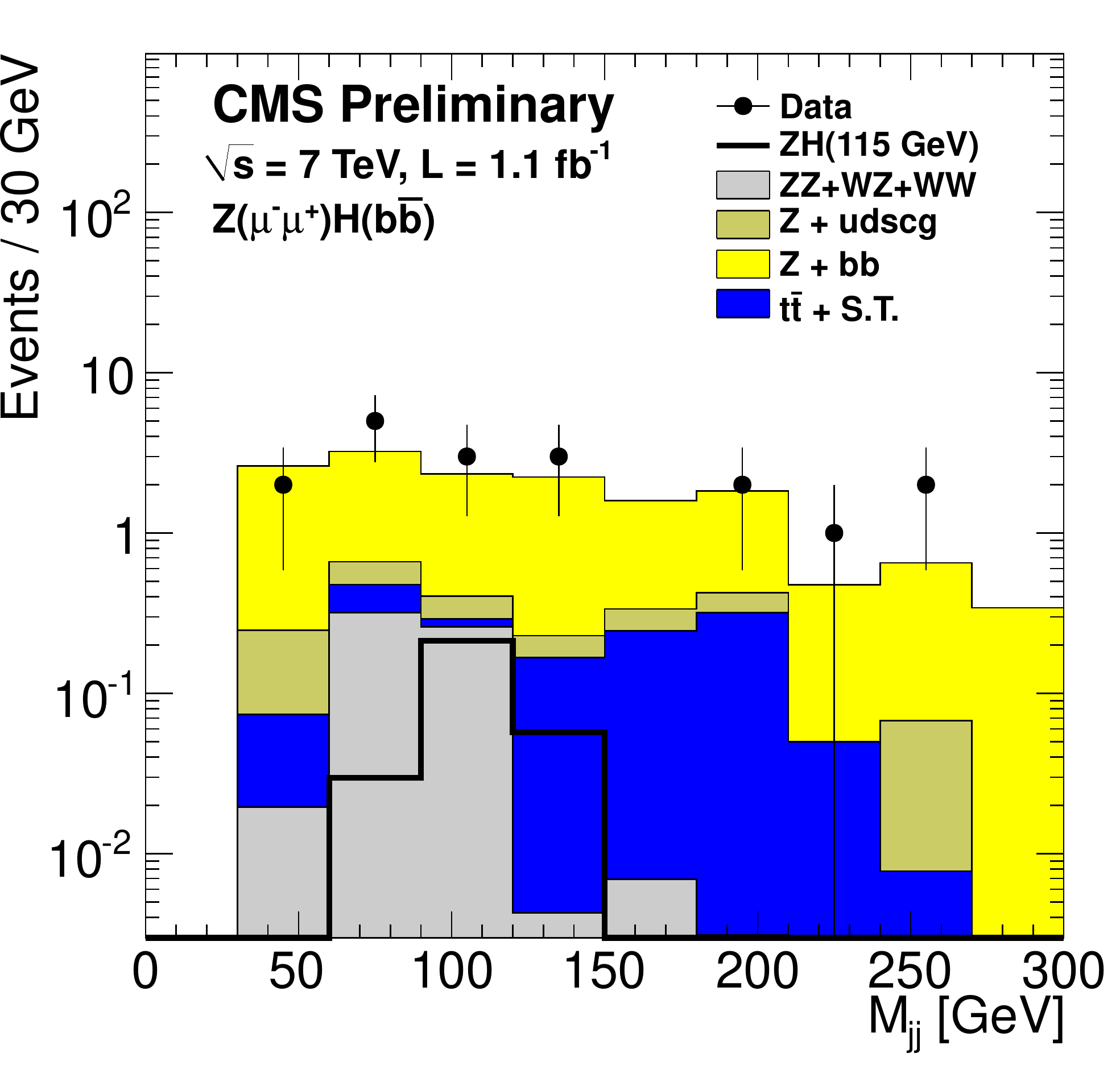}}
  \resizebox{0.62\columnwidth}{!}{  \includegraphics{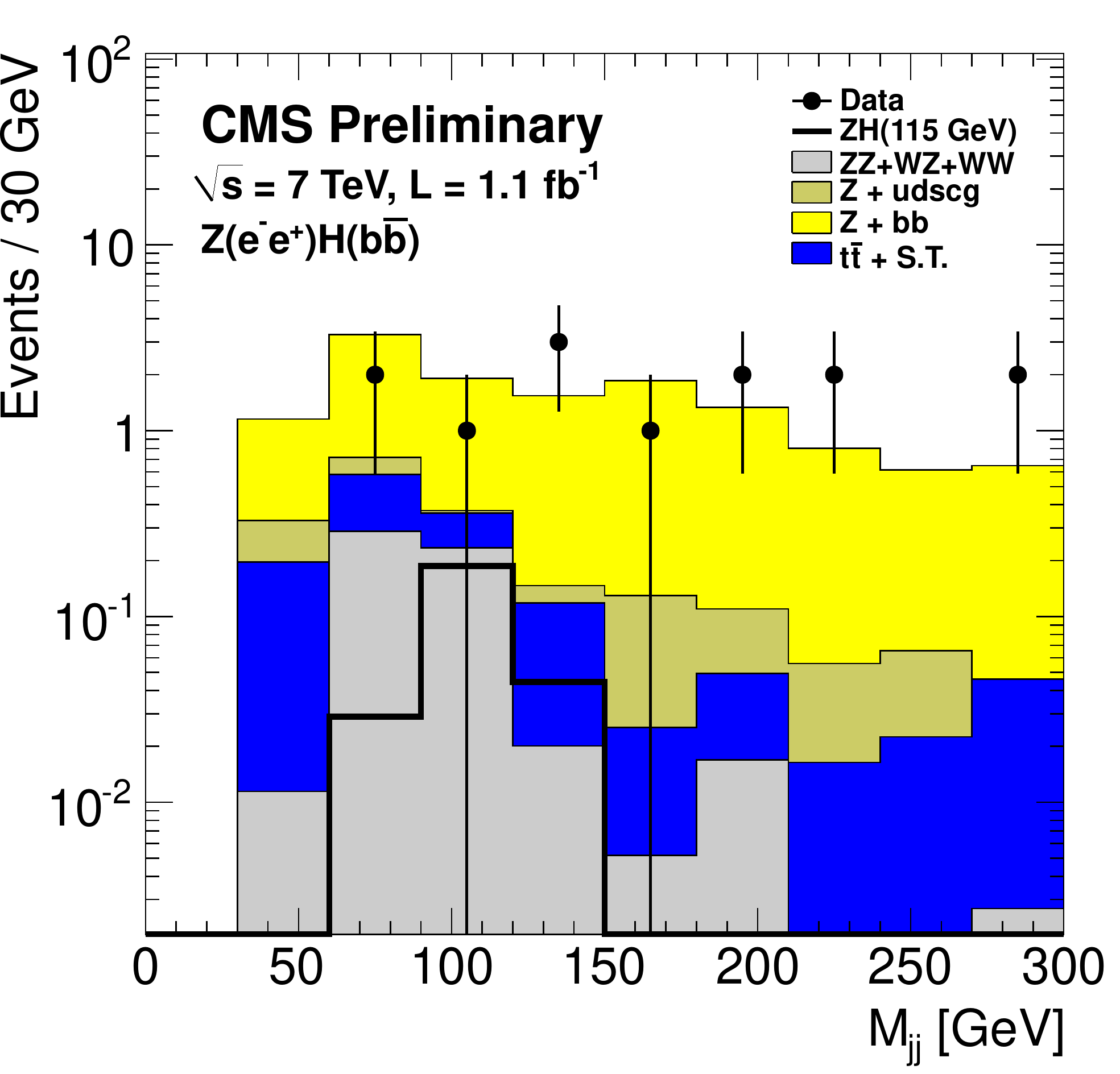}}
  \resizebox{0.62\columnwidth}{!}{  \includegraphics{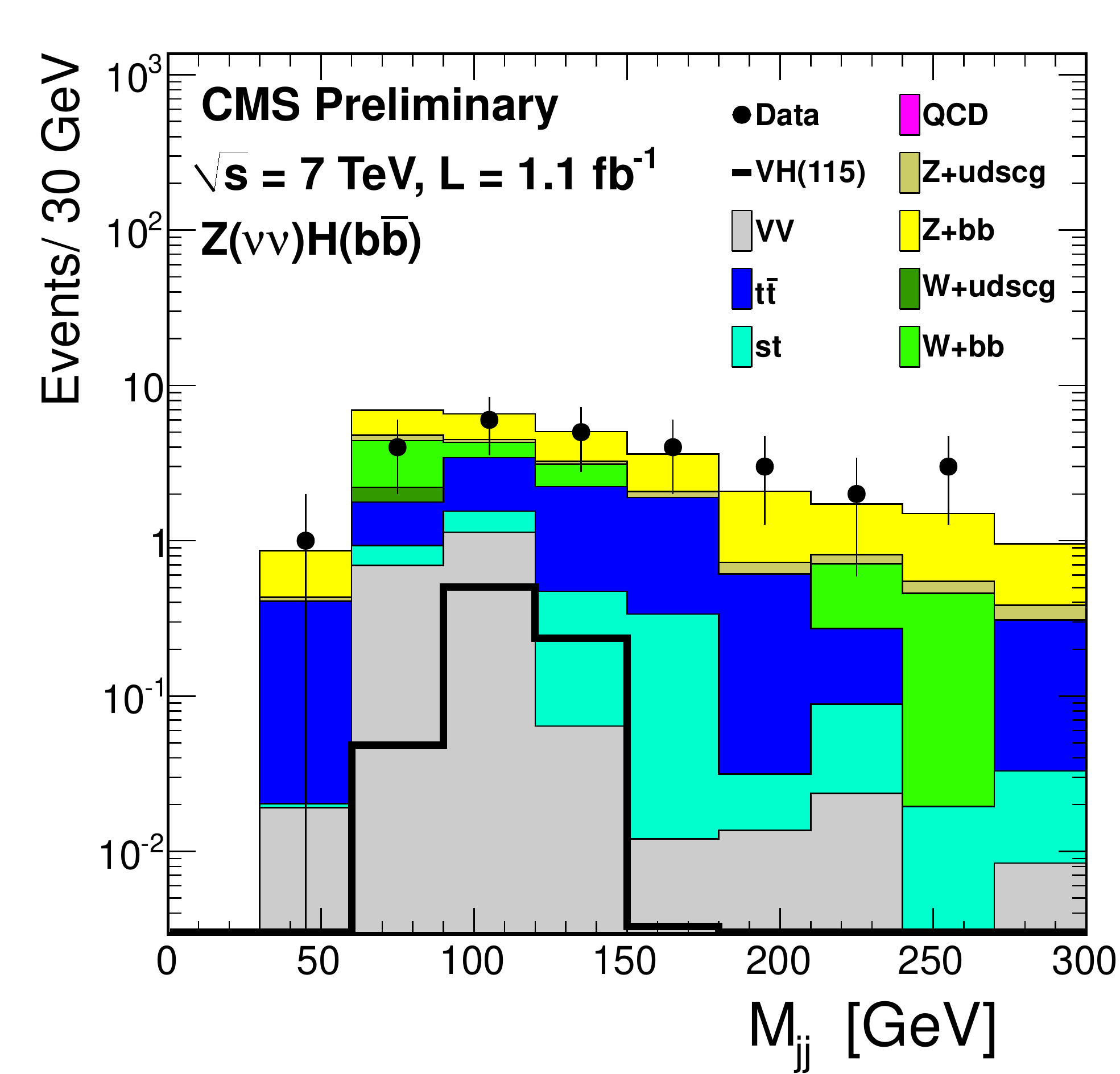}}
  \caption{Distributions of dijet invariant mass after all \Mbb\  selection criteria have been applied in (from top to  bottom: \WmnH, \WenH, \ZmmH, \ZeeH, \ZnnH).}
\label{fig:MassCC}
\end{center}
\end{figure}

\begin{figure}[htbp]
 \begin{center}
 \resizebox{0.62\columnwidth}{!}{  \includegraphics{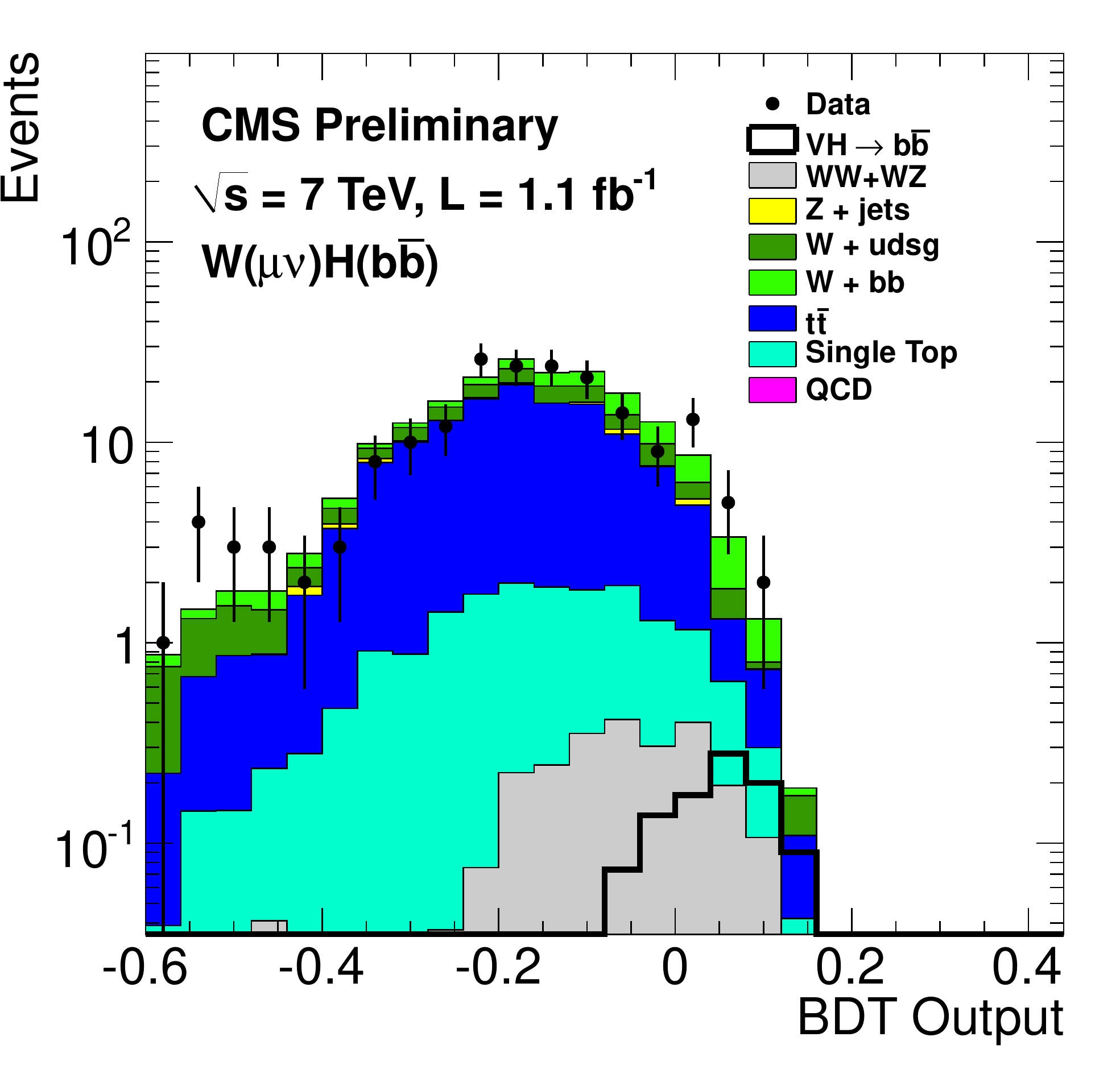}}
 \resizebox{0.62\columnwidth}{!}{  \includegraphics{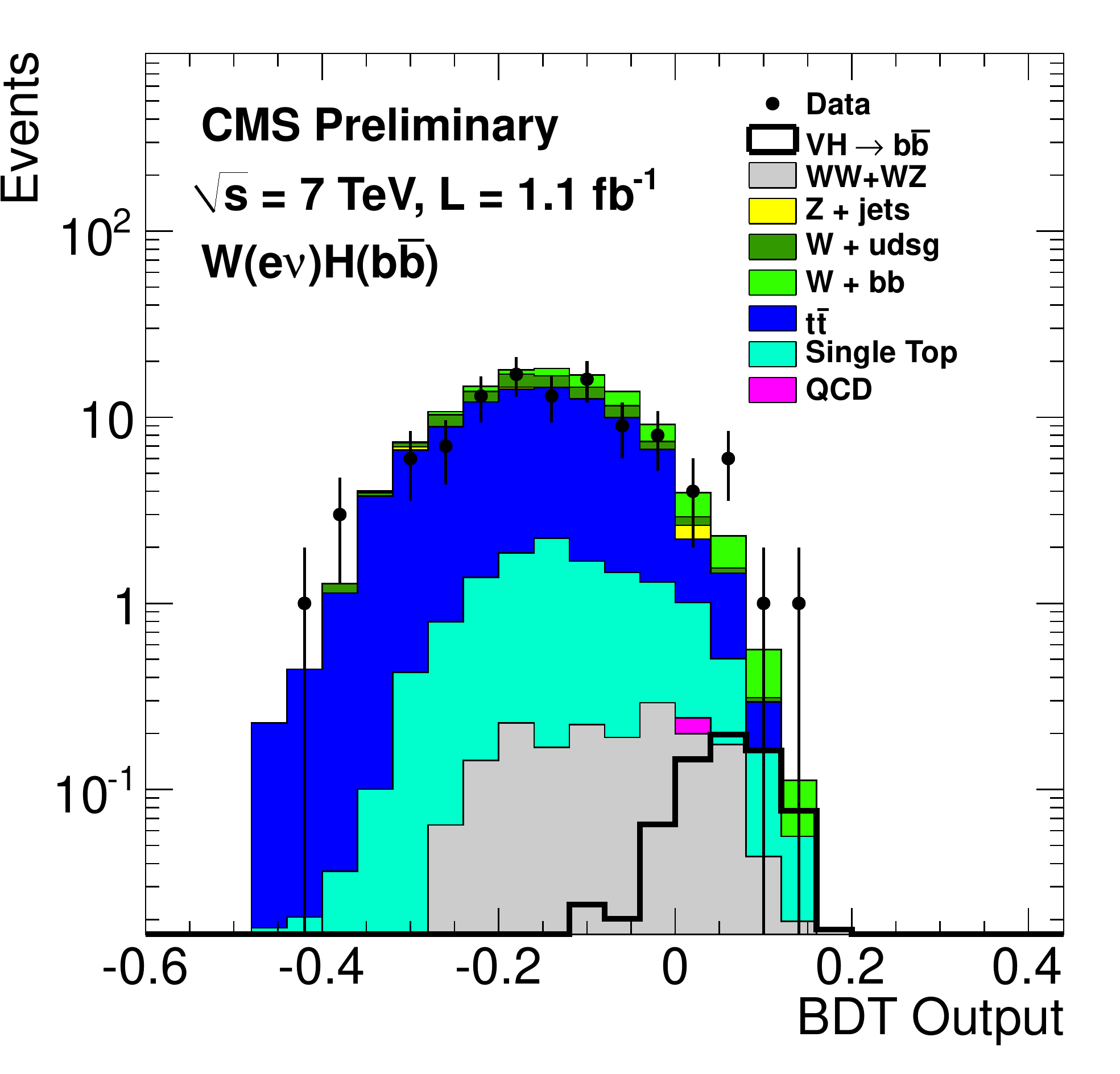}}
 \resizebox{0.62\columnwidth}{!}{  \includegraphics{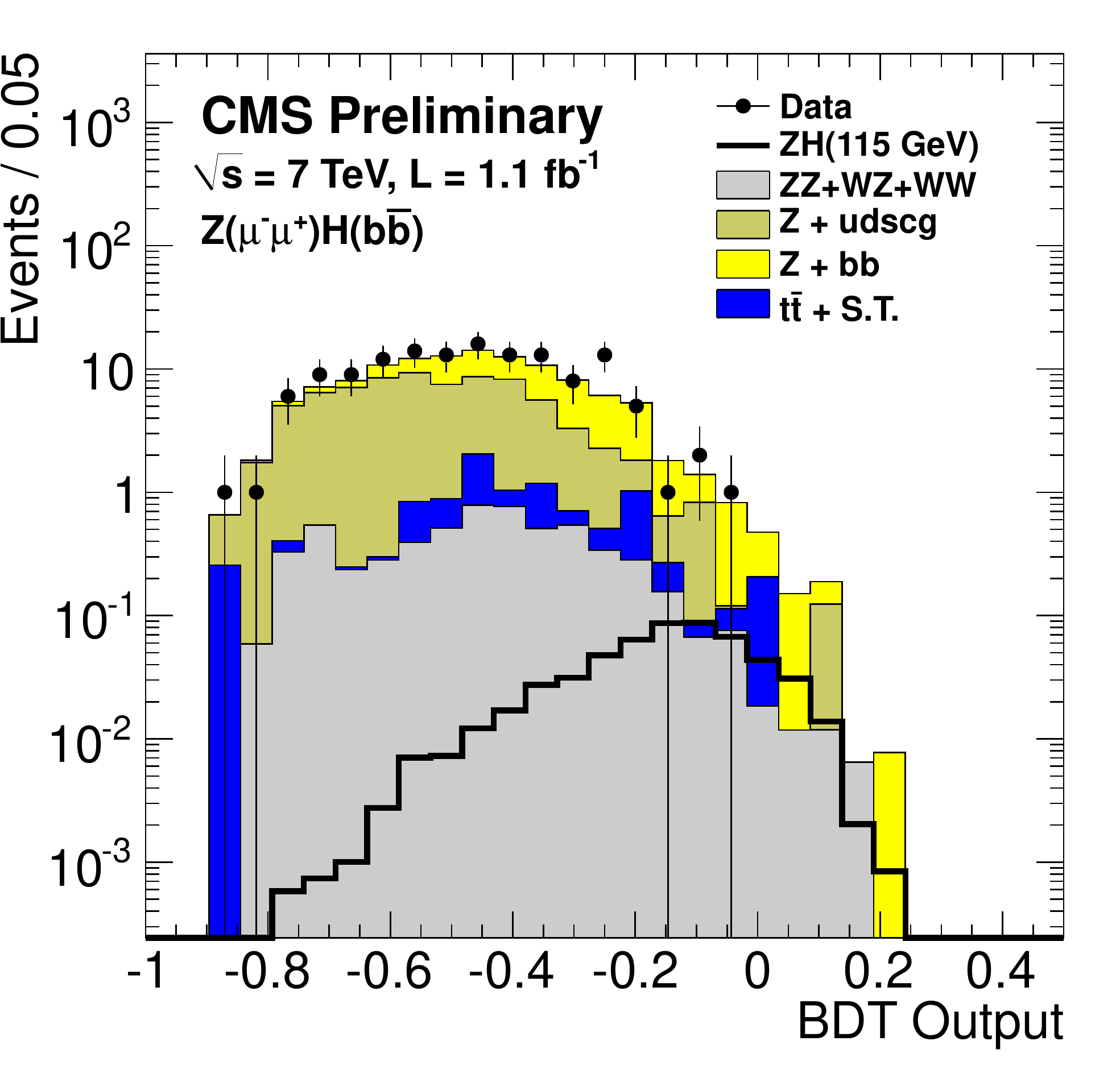}}
 \resizebox{0.62\columnwidth}{!}{  \includegraphics{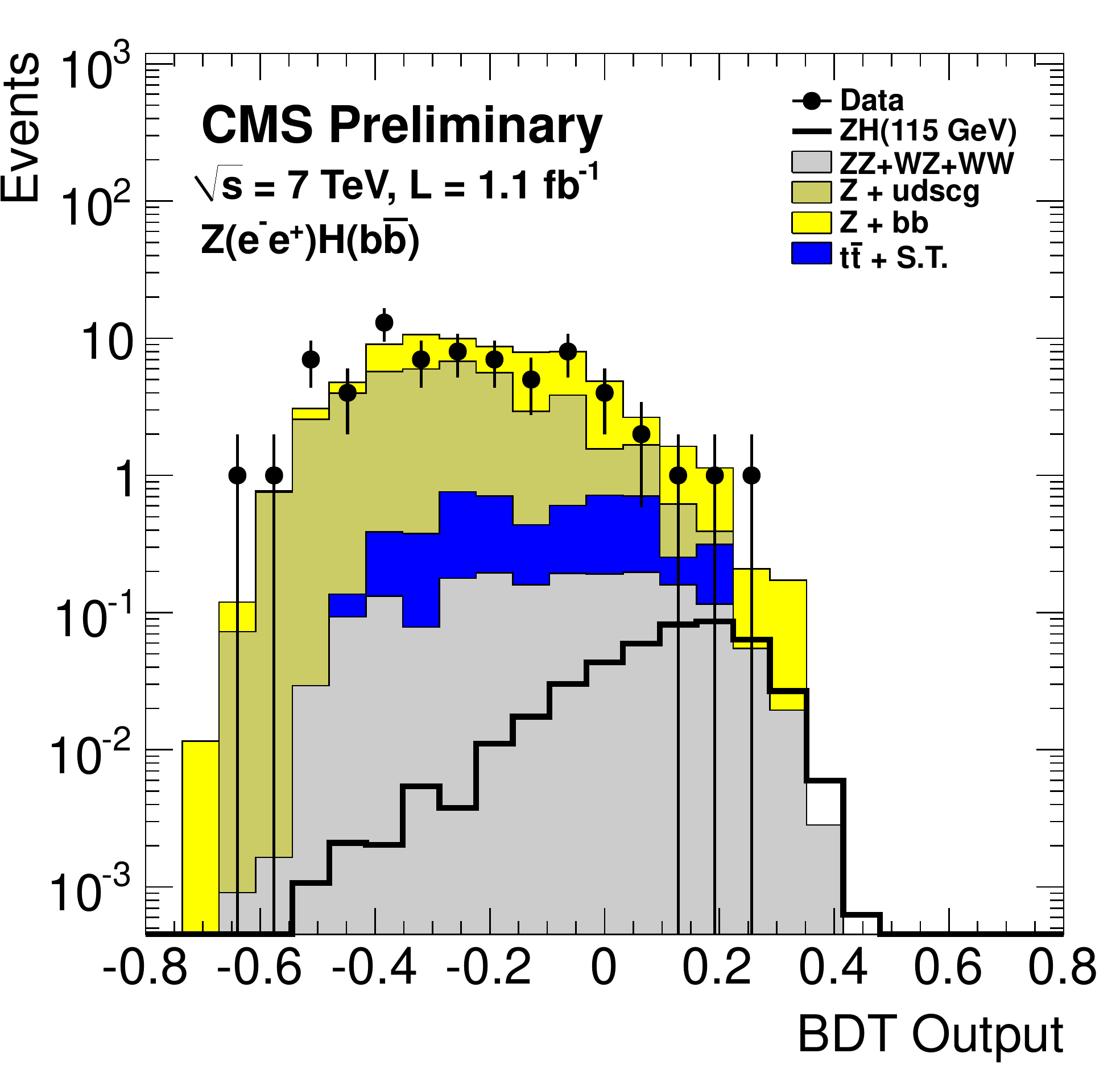}}
 \resizebox{0.62\columnwidth}{!}{  \includegraphics{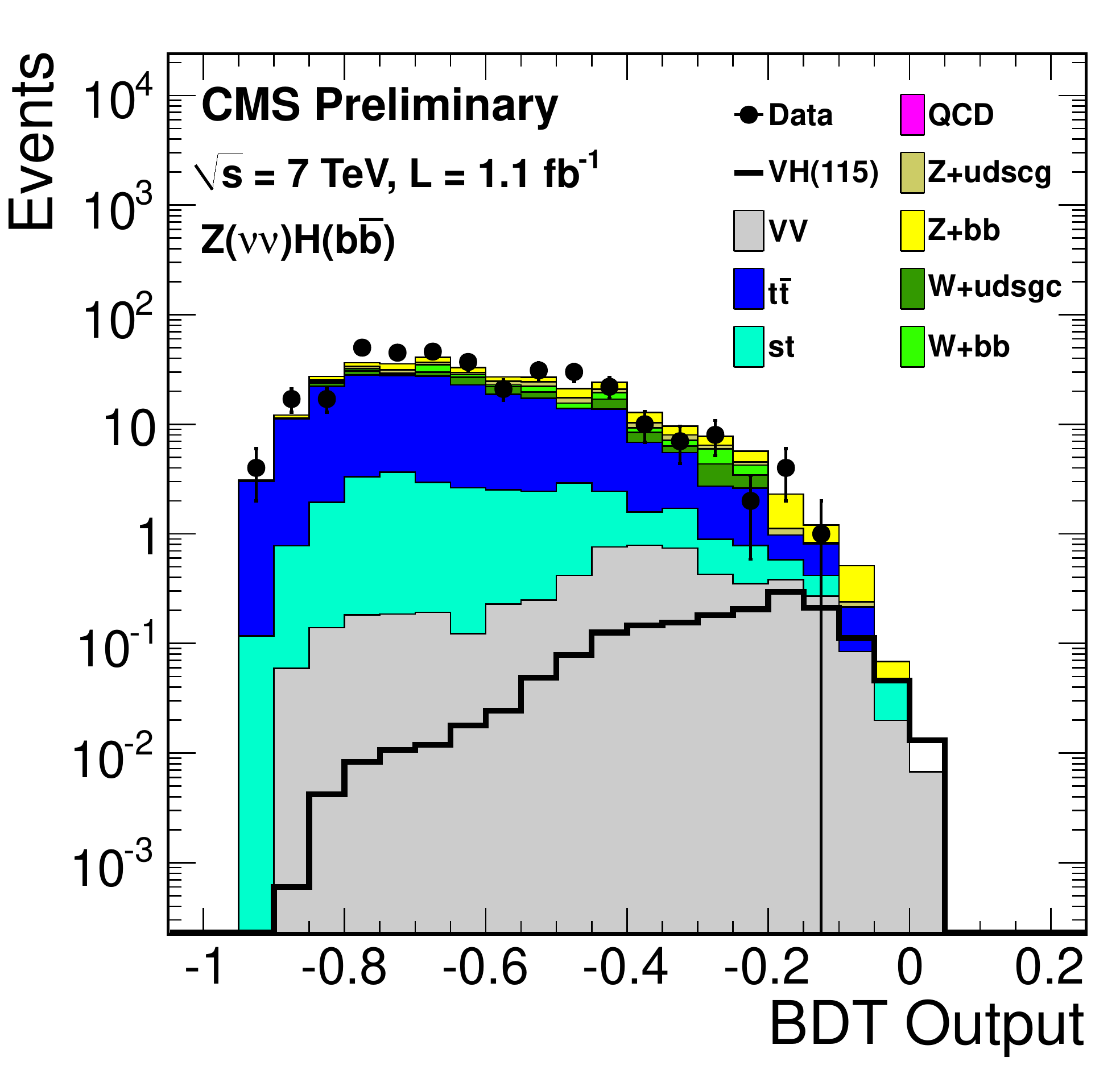}}
    \caption{Distributions of \BDT\ output for  data (points with
      errors) and all backgrounds, from top to bottom: \WmnH, \WenH, \ZmmH, \ZeeH  and \ZnnH.}
    \label{fig:BDTdata}
  \end{center}
\end{figure}

The final predicted number of events in the signal regions of the
\BDT\ and \Mbb\ analyses are determined with a mix of data-driven
estimates based on the control regions, 
and expectations from simulation.  We summarize the
final signal and background estimates in both sets of analyses, including
the systematic uncertainties summarized in the previous section, and the
expected and observed upper limits using 1.1~fb$^{-1}$ of integrated luminosity.
We report in tables \ref{tab:CCyields} and \ref{tab:BDTyields} and
figures \ref{fig:MassCC} and \ref{fig:BDTdata}   the results for a single
mass point, 115 GeV. While the final limits plots include mass points from 110 to 135 GeV.

\begin{table}[htbp]
\caption{Predicted backgrounds, signal 
yields with total uncertainty, and the observed number of events for 115 mass 
point for the 5 channels \Mjj\ analysis. We report also the sliding
windows on \Mjj\ for the 115 mass point search.}
\label{tab:CCyields}
\begin{center}
\scalebox{0.65}{
\begin{tabular}{ccccccc} \hline\hline
Process     & $\WmnH$       & $\WenH$       &  $\ZmmH$        &
$\ZeeH$ & $\ZnnH$\\ \hline  
\Mjj\ cut & $100$--$130$ & $100$--$130$  & $95$--$125$ &$95$--$125$& $100$--$130$ \\ \hline 
\Wudscg     & $0.081\pm 0.038$& $0.01 \pm 0.004$ & - & - & $0.023\pm 0.007$ \\
\Wbb        & $0.829\pm 0.221$& $0.344\pm 0.093$ & - &  - & $0.310\pm 0.084$\\
\Zudscg  &- & - & $0.110\pm 0.065$ &   $0.006\pm 0.003$ & $0.180\pm 0.039$  \\
\Zbb      & $0.184\pm 0.131$& $0.204\pm 0.146$ & $2.050\pm 0.396$ & $1.545\pm 0.254$ & $1.890\pm 0.578$\\
\ttbar      & $1.109\pm 0.287$& $0.543\pm 0.136$ & $0.090\pm 0.036$ &  $0.133\pm 0.073$ & $1.470\pm 0.504$ \\ 
ST          & $0.24 \pm 0.105$& $0.122\pm 0.049$ & $0.090 \pm 0.036$ & $0.009\pm 0.007$ &  $0.410\pm 0.156$  \\\hline
VV          & $0.153\pm 0.064$& $0.065\pm 0.026$ &  $0.160\pm 0.064$ & $0.189\pm 0.074$ & $0.460\pm 0.174$  \\
\Bexp       & $2.596\pm 0.449$& $1.288\pm 0.242$ & $2.410\pm 0.358$ &$1.883\pm 0.232$ &  $4.793\pm 0.938$  \\\hline
WH          & $0.354\pm 0.099$& $0.296\pm 0.083$ &  - & - & $0.091\pm 0.012$\\
ZH          & $0.006\pm 0.002$& $0.002\pm 0.001$ & $0.195\pm 0.051$ &$0.193\pm 0.050$ & $0.502\pm 0.104$ \\ \hline           
\Nobs       & $4$   	    & $4$ &      $3$  & $2$     &5  \\  
\hline\hline
\end{tabular} 
}
\end{center}
\end{table}

\begin{table}[htbp]
\caption{Predicted backgrounds, signal 
yields with total uncertainty, and the observed number of events for 115 mass 
point for the 5 channels \BDT\ analysis. We report also the \BDT\ cut we
choose for the search.}
\label{tab:BDTyields}
\begin{center}
\scalebox{0.65}{
\begin{tabular}{ccccccc} \hline\hline
Process     & $\WmnH$       & $\WenH$       &  $\ZmmH$        &
$\ZeeH$ & $\ZnnH$\\ \hline  
BDT        & $>0.050$  & $>0.040$  & $>-0.145$  & $>0.160$  & $>-0.175$\\ \hline
\Wudscg     & $0.667\pm 0.192 $& $0.155\pm 0.063$ & - & - & -  \\
\Wbb        & $2.035\pm 0.533$& $1.378\pm 0.374$ & - &  - & $0.359\pm 0.097$\\
\Zudscg  &- & - & $0.110\pm 0.065$ &   $0.077\pm 0.036$ &  \\
\Zbb      & $0.006\pm 0.006$& $0.198\pm 0.141$ & $2.858\pm 0.645$ & $0.904\pm 0.189$ & $1.108\pm 0.339$\\
\ttbar      & $1.173\pm 0.304$& $1.254\pm 0.315$ & $0.296\pm 0.118$ &  $0.179\pm 0.133$ & $0.297\pm 0.077$ \\ 
ST          & $0.653\pm 0.261$& $0.653\pm 0.261$ & $0.060\pm 0.014$ & $-$ &  $0.453\pm 0.172$ & \\\hline
VV          & $0.355\pm 0.140$& $0.292\pm 0.117$ &  $0.334\pm 0.030$ &
$0.195\pm 0.078$ & $0.571\pm 0.216$  \\
\Bexp       & $4.889\pm 0.806$& $3.930\pm 0.658$ & $4.773\pm 0.641$ &$1.354\pm 0.240$ &  $2.901\pm 0.572$ \\\hline
WH          & $0.587\pm 0.164$& $0.477\pm 0.134$ &  - & - & $0.507\pm 0.112$\\
ZH          & $0.011\pm 0.003$& $0.004\pm 0.001$ & $0.328\pm 0.085$ &$0.183\pm 0.048$ & $0.049\pm 0.007$ \\ \hline           
\Nobs       & $7$   	    & $9$ &      $4$  & $2$     &$1$&  \\  
\hline\hline
\end{tabular} 
}
\end{center}
\end{table}

\section{Upper Limits}
Preliminary  $95\%$ C.L. upper limits on the Higgs production
cross section in the VH mode with \HBB\ were obtained from both
the \BDT\ and \Mjj\ analyses for a dataset corresponding to an integrated
luminosity of  $ 1.1$~fb$^{-1}$. For the expected and observed limits,
and the 1- and 2-$\sigma$ bands, the CLs method currently recommended by the LHC Higgs Combination Group was employed~\cite{HIG-11-011}. 

\begin{figure}[tbp]
  \begin{center}
%    \includegraphics[width=0.49\textwidth]{figures/BDTlimitCLs.pdf}
%    \includegraphics[width=0.49\textwidth]{figures/CClimitCLs.pdf}
%\begin{minipage} 
\resizebox{0.75
  \columnwidth}{!}{{ \includegraphics{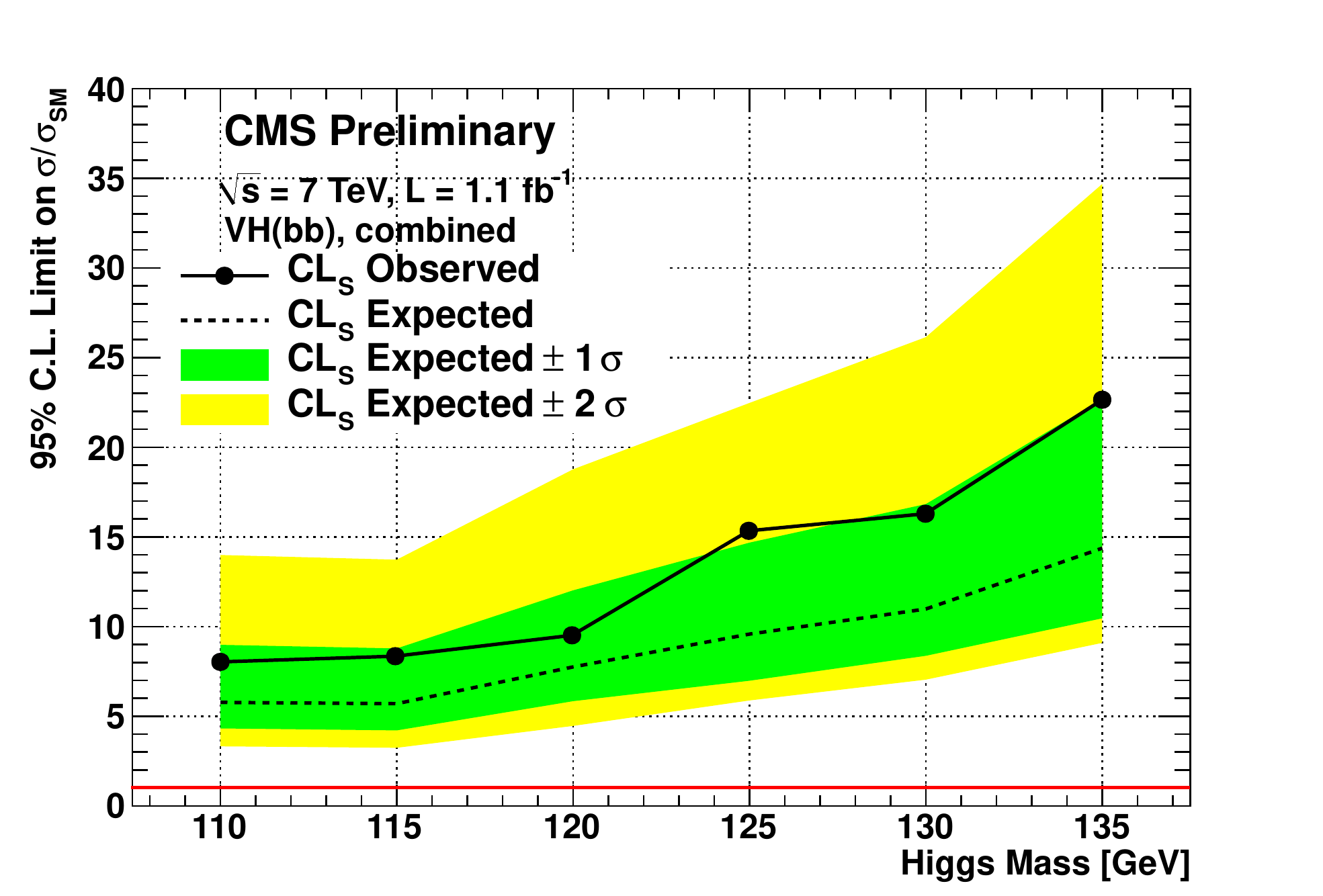}}}
%\end{minipage}
%\begin{minipage}
 %{4cm}
{ \resizebox{0.75     \columnwidth}{!}{\includegraphics{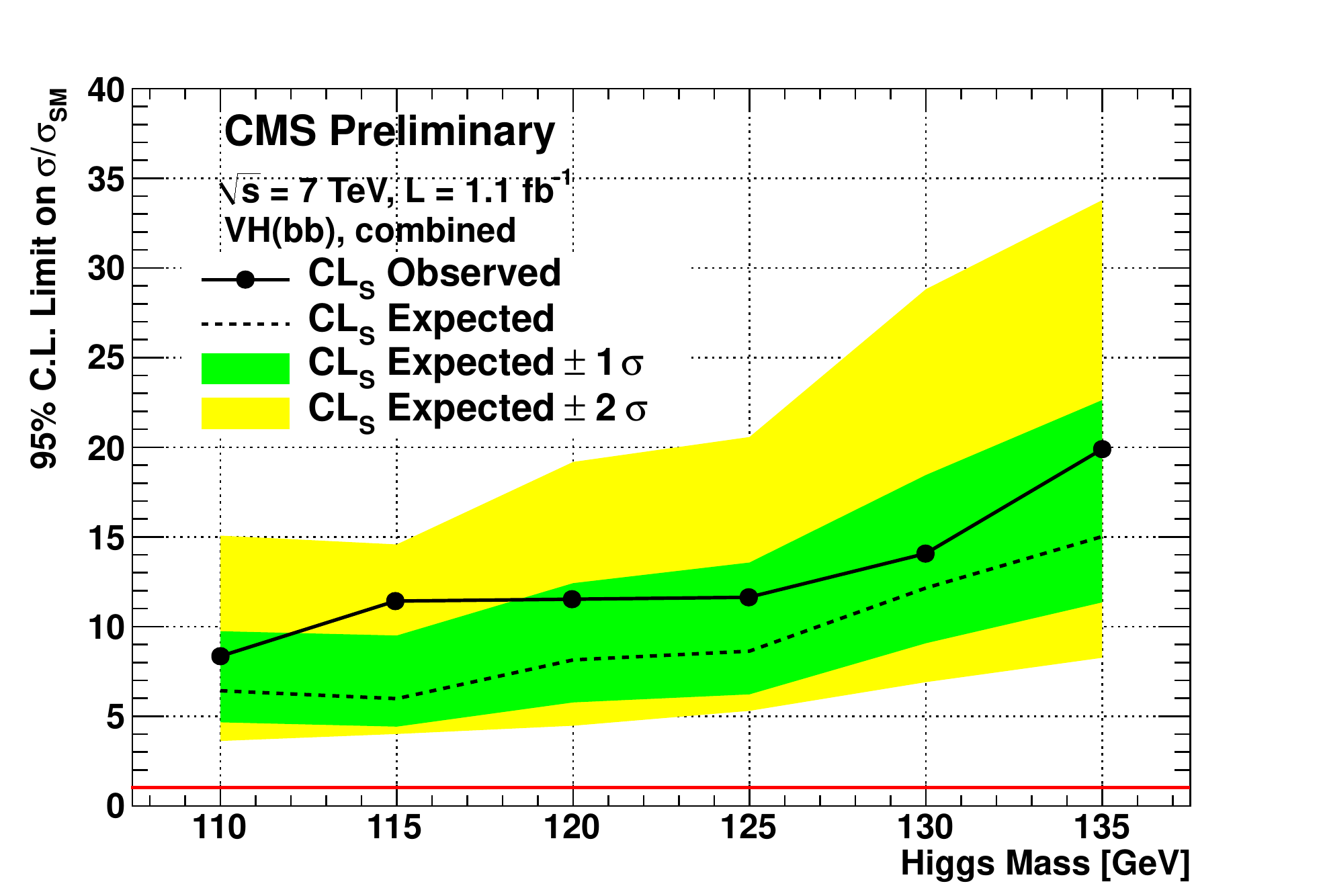}}}
%\end{minipage}
    \caption{Expected and observed $95\%$ C.L. combined upper limits on the ratio of
    VHbb production for the BDT (top) and \Mjj\ (bottom) analyses.  The median expected
    limit and the 1- and 2-$\sigma$ bands are obtained with the LHC CLs method as
    implemented in LandS, as are the observed limits at each mass point.}
    \label{fig:Limits}
  \end{center}
\end{figure}

The results of the five \BDT\ analyses are combined to produce limits on 
Higgs production in the $b\bar{b}$\ channel for the assumed masses:
$110-135$~GeV.  The identical procedure was applied to the results of the 
\Mbb\ analysis. Table~\ref{tab:Limits}
summarizes the resulting, expected and observed, upper $95\%$
C.L. cross section limits,
with respect to the standard model cross section, for each of the mass 
points for the \BDT\ and \Mbb\ analyses. The results are displayed separately
in Fig.~\ref{fig:Limits}.  The primary result is the one from the \BDT\ analysis.

%The expected Bayesian limits are compared to the CLs 
%results in Fig.~\ref{fig:BayesLimits}.

\begin{table}[tbp]
\caption{Expected and observed $95\%$ CL upper limits on the 
production of a SM Higgs boson in association with W and Z bosons
and decaying to b quarks relative to the expected cross section.
Limits are listed separately for the \BDT\ and \Mbb\ analyses.}
\label{tab:Limits}
\begin{center}
\scalebox{0.7}{
\begin{tabular}{ccccc} \hline\hline
$M_H$ (GeV) & \BDT\ Expected & \BDT\ Observed & \Mjj\ Expected & \Mjj\ Observed \\  \hline
 110      & $5.8$        & $8.0$        & $6.4$          & $8.2$          \\
 115      & $5.7$        & $8.3$        & $6.0$          & $11.3$         \\
 120      & $7.7$        & $9.5$        & $8.1$          & $11.4$         \\
 125      & $9.6$        & $15.3$       & $8.6$          & $11.6$         \\
 130      & $11.0$       & $16.3$       & $12.1$         & $14.0$         \\
 135      & $14.4$       & $22.5$       & $15.0$         & $19.9$         \\
\hline\hline
\end{tabular}
}
\end{center}
\end{table}

\end{document}